\documentstyle[12pt,aasms,psfig]{article}

\def \etal {et al.\thinspace}
\newcommand{\en}{\mbox{$n$}\ }
\newcommand{\el}{\mbox{$\ell$}\ }

\newcommand{\ang}{\mbox{$\AA$}\ }

\received{ 2000}
\journalid{337}{15 January 1989}
\articleid{11}{14}

\slugcomment{Submitted to Astrophys. J. {\it Suppl.}, 2000}

\begin{document}

\title{Electron-Ion Recombination Rate Coefficients and Photoionization
Cross Sections for Astrophysically Abundant Elements. VII.
Relativistic calculations for O VI and O VII for UV and X-ray modeling}

\author{Sultana N. Nahar and Anil K. Pradhan}
\affil{Department of Astronomy, The Ohio State University,
Columbus, OH 43210}

\begin{abstract}

Aimed at ionization balance and spectral analysis of UV and X-ray sources,
we present
self-consistent sets of photoionization cross sections, recombination
cross sections, and rate coefficients for Li-like O~VI and He-like O~VII.
Relativistic fine structure is considered through the
Breit-Pauli R-matrix (BPRM) method in the
close coupling approximation, implementing the unified treatment for
total electron-ion recombination subsuming both radiative and di-electronic
recombination processes. Self-consistency is ensured
by using an identical wavefunction expansion for the inverse processes
of photoionization and photo-recombination. Radiation damping of resonances,
important for H-like and He-like core ions, is included.
Compared to previous LS coupling results without radiative decay of
low-\en \en $\leq$ 10) resonances, the
presents results show significant reduction in O VI recombination rates
at high temperatures. In addition to the total rates, level-specific
photoionization cross sections and recombination rates are presented for
all fine structure levels \en (\el SLJ) up to \en $\leq 10$, to enable accurate
computation of recombination-cascade matrices and spectral formation of
prominent UV and X-ray lines such as the 1032,1038 \ang doublet of O~VI,
 and the `triplet' forbidden, intercombination, and resonance X-ray
lines of O~VII at 22.1, 21.8, and 21.6 \ang respectively. Altogether,
atomic parameters for 98 levels of O~VI and 116 fine structure levels of
O~VII are theoretically computed. These data should provide a reasonably
complete set of photoionization and recombination rates
in collisional or radiative equilibrium.

\end{abstract}

\keywords{atomic data --- atomic processes ---  photoionization,
dielectronic recombination, unified electron-ion recombination ---
X-rays: general --- line:formation}

\section{INTRODUCTION}

 Li-like and He-like Oxygen are among the most important atomic species
in UV and X-ray plasma diagnostics of astrophysical sources. For
example, there is considerable current interest in O~VI and O~VII as
a possible reservoir of `missing baryons' in the warm/hot
intergalactic medium in the present universe (Cen \etal 2001, Fang and
Canizares 2000, Yoshikawa \etal 2003). In addition to global ionization
balance models that require O~VI/O~VII photoionization and recombination
rates, specific spectroscopic diagnostics of both of these ions are
related in an interesting but intricate manner. Outflows observed from
the `warm absorber' in active galactic nuclei exhibit absorption/emission
lines in the X-ray from KLL resonant photoabsorption in O~VI, and
excitation/recombination of O~VII lines. This phenomenon was
investigated theoretically
by Pradhan (2000) who pointed out that the observed spectra of the well
known AGN NGC 5548 shows these features (Kaastra \etal 2000).
Recently Arav \etal (2003) have further pursued this suggestion to infer
large discrepancies in the UV and X-ray O~VI column densities in the outflow
from NGC 5548. Ionization balance and spectral formation of  O~VI and
O~VII are crucial factors in all such studies.

The inverse atomic processes of photoionization and electron-ion
photorecombination of Li-like and He-like ions are also of
particular interest in X-ray astronomy to analyze new observations
by space based observatories such as the Chandra X-ray Observatory and
and XMM-Newton, at photon energies and temperatures prevalent in
high-temperature sources such as AGB, supernova remnants, hot stellar
coronae etc. (e.g. Canizares \etal 2000,  in
Proceedings of the NASA workshop on {\it Atomic data needs in X-ray
astronomy} 2000). X-ray emission in the
K$\alpha$ complex comprising of the `triplet' feature of resonance (w),
intercombination (x,y), and forbidden (z) lines in He-like ions,
corresponding to the 4 transitions from the \en = 2 levels to the ground
level $1s^2 \ (^1S_0) \longleftarrow 1s2p (^1P^o_1), 1s2p (^3P^o_{2,1}),
1s2s (^3S_1)$ respectively, yields valuable spectral diagnostics
of temperature, density, ionization balance, and abundances in the plasma
source.
Calculations of the recombination-cascade contributions
for these important lines requires accurate
atomic parameters for fine structure levels up to fairly high \en
levels, as presented in this report.

 In the Present work we report new results for self-consistent
data sets for photoionization and recombination for (e + O V) $\rightarrow$
O VI and (e + O VI) $\rightarrow$ O VII, using a
unified theoretical method that takes account of both radiative and
di-electronic recombination processes (RR and DR).
Although treated separately in
other previous treatments, both recombination processes, RR and
DR, are inseparable in nature. A unified treatment is therefore not only
theoretically and computationally more
accurate, it is also more practically convenient for applications
since a single recombination rate coefficent takes account of both RR
and DR in an ab initio manner.
For highly charged ions it is important to consider relativistic fine
structure explicitly in the theoretical formulation, in addition to the
electron correlation effects. Previous results for O~VI
(Nahar 1998, 1999) were obtained in a non-relativistic LS coupling
approximation; relativistic effects were partially included for O~VII.
 To further improve the accuracy, as well as to provide extensive
sets of data for astrophysical models, we have re-calculated all
photoionization cross sections, electron-ion recombination cross
sections, and rate coefficients using the relativistic Breit-Pauli R-matrix
(BPRM) method. Similar
calculations have previously been reported for He- and Li-like ions
C~V and C~IV (Nahar \etal 2000), and Fe XXV and Fe XXV (Nahar et al. 2001).

\section{THEORY}

The unified method of electron-ion recombination (Nahar and Pradhan
1992) yields self-consistent sets of atomic paramters for
photoionization and recombination for atoms and ions (e.g. Nahar and
Pradhan 1997, Paper I of this series). Relativistic extension using the
BPRM method is described in Zhang and Pradhan (1997), Pradhan and Zhang
(1997) and Zhang \etal (1999). Photorecombination of an incident
electron with the target ion may occur through (i) non-resonant,
background continuum, or radiative recombination (RR),
\begin{equation}
e + X^{++} \rightarrow  h\nu + X^+,
\end{equation}
which is the inverse process of direct photoionization, or (ii) through
a two-step recombination process via autoionizing resonances, i.e.
dielectronic recombination (DR):
\begin{equation}
e + X^{++} \rightarrow (X^+)^{**}  \rightarrow  \left\{
\begin{array}{c} (i) \ e + X^{++} \\ (ii) \  h\nu + X^+ \end{array}
\right. ,
\end{equation}
where the quasi-bound doubly-excited autoionizing state may lead either
to (i) autoionization, a radiation-less transition to a lower target
state with the electron going into a continuum, or (ii) radiative
stabilization to a recombined bound state via decay of the ion core
(usually to the ground state) with the electron captured.

 The extension of the R-matrix method to electron-ion recombination
calculations entails close coupling calculations for photoionization
and electron-ion scattering. Identical eigenfunction expansion for the
target (core) ion is employed for both processes, enabling inherently
self-consistent results in an ab initio manner for a given ion.
We consider photoionization from, and recombination into, the
infinity of levels of the (e~+~ion) system. These are divided into
two groups of bound levels: (A) with $\nu \leq \nu_o$ and  all
possible fine structure $J\pi$ symmetries, and (B)
$ \nu_o < \nu \leq \infty $; where $\nu$ is the effective quantum number
relative to the target threshold(s). Photoionization and recombination
calculations are carried out in detail for all group A levels, while
group (B) levels are treated through quantum defect theory of DR within
close the coupling approximation (Bell and Seaton 1985, Nahar and Pradhan
1994). A generally valid approximation made in recombination to group
(B) levels is that the background contribution is negligible, and DR is
is the dominant process in the region below the threshold of convergenece
for high-\en resonances.

In the close coupling (CC) approximation the target ion (core) is
represented by an $N$-electron system. The total wavefunction, $\Psi(E)$,
of the ($N$+1) electron-ion system of symmetry $J\pi$ is
represented in terms of an expansion of target eigenfunctions as:
\begin{equation}
\Psi(E) = A \sum_{i} \chi_{i}\theta_{i} + \sum_{j} c_{j} \Phi_{j},
\end{equation}
where $\chi_{i}$ is the target wavefunction in a specific level
$J_i\pi_i$ and $\theta_{i}$ is the wavefunction for the
($N$+1)-th electron in a channel labeled as
$S_iL_i(J_i)\pi_ik_{i}^{2}\ell_i(\ J\pi))$; $k_{i}^{2}$
being its incident kinetic energy. $\Phi_j$'s are the correlation
functions of the ($N$+1)-electron system that account for short range
correlation and the orthogonality between the continuum and the bound
orbitals.
In relativistic BPRM calculations the set of ${SL\pi}$ are recoupled
for $J\pi$ levels of (e + ion)-system, followed by diagonalisation of
the Hamiltonian,
\begin{equation}
H^{BP}_{N+1}\mit\Psi = E\mit\Psi,
\end{equation}
where the BP Hamiltonian is
\begin{equation}
H_{N+1}^{\rm BP}=H_{N+1}^{NR}+H_{N+1}^{\rm mass} + H_{N+1}^{\rm Dar}
+ H_{N+1}^{\rm so}.
\end{equation}
The first term, $H_{N+1}^{NR}$, is the nonrelativistic Hamiltonian,
\begin{equation}
H_{N+1}^{NR} = \sum_{i=1}\sp{N+1}\left\{-\nabla_i\sp 2 - \frac{2Z}{r_i}
        + \sum_{j>i}\sp{N+1} \frac{2}{r_{ij}}\right\},
\end{equation}
and the additional one-body terms are the mass correction term,
$H^{\rm mass} = -{\alpha^2\over 4}\sum_i{p_i^4}$, the Darwin term,
$H^{\rm Dar} = {Z\alpha^2 \over 4}\sum_i{\nabla^2({1\over r_i})}$. and
the spin-orbit interaction term, $H^{\rm so}= Z\alpha^2 \sum_i{1\over r_i^3}
{\bf l_i.s_i}.$, respectively.

The positive and negative energy states (Eq. 4) define continuum or
bound (e~+~ion) states, such that, $E = k^2 >$ 0 for  continuum
(scattering) channels and $E = - \frac{z^2}{\nu^2} <$ 0 for bound states,
where $\nu$ is the effective quantum number relative to the core level.
If $E <$ 0 then all continuum channels are `closed' and the solutions
represent bound states. The matrix element for the bound-free transition,
$<\Psi_B || {\bf D} || \Psi_{F}>$, can be obtained from the continuum
wavefunction ($\Psi_{F}$) and the bound wavefunction ($<\Psi_B$).
{\bf D} is the dipole operator, ${\bf D}_L = \sum_i{r_i}$, in length
form and ${\bf D}_V = -2\sum_i{\Delta_i}$ in velocity form, the sum being
the number of electrons.

Photoinization cross sections of all possible states with
$n \leq n_{\rm max}\sim 10$ -- are obtained as in the CC approximation
as in the Opacity Project (Seaton 1987, {\it The Opacity Project} 1995,
1996), extended to include the relativistic effects under the Iron
Project (Hummer et al. 1993). The photoionization cross section is
obtained as
\begin{equation}
\sigma_{PI} = {1\over g}{4\pi^2\over 3c}\omega{\bf S},
\end{equation}
where $g$ is the statistical weight factor of the bound state and ${\bf S}$
is the dipole line strength, ${\bf S}~=~|<\Psi_B || {\bf D} || \Psi_F >|^2$.
For highly charged H- and the He-like ions, the probability of radiation
decay of an autoionizing state is usually comparable to that of
autoionization as discussed in Nahar \etal (2000). With strong dipole
allowed $ 2p \longrightarrow 1s$ and $ 1s2p \ (^1P^o_1) \longrightarrow
1s^2 \ (^1S_0)$ transitions autoionizing resonances are radiatively damped
to a significant extent.
The radiative damping effect of all near-threshold resonances, up to
$\nu \leq 10$, is included using a resonance fitting procedure
(Sakimoto et al. 1990, Pradhan and Zhang 1997, Zhang \etal 1999).

The photo-recombination cross section, $\sigma_{\rm RC}$, is related to
photoionization cross section, $\sigma_{\rm PI}$, through principle of
detailed balance (Milne relation) as
\begin{equation}
\sigma_{\rm RC}(\epsilon) =
{\alpha^2 \over 4} {g_i\over g_j}{(\epsilon + I)^2\over \epsilon}
\sigma_{\rm PI}
\end{equation}
\noindent
in Rydberg units; $\alpha$ is the fine structure constant, $\epsilon$ is
the photoelectron energy, and $I$ is the ionization potential.
$\sigma_{\rm RC}$ are computed from the photoionization cross sections at a
sufficiently large number of energies to delineate the non-resonant
background  and the autoionizing resonances, thereby representing both
radiative and the dielectronic recombination (RR and DR) processes.
In the unified treatment the photoionization cross sections,
$\sigma_{\rm PI}$, of a large number of low-$n$ bound states -- all
possible states with $n \leq n_{\rm max}\sim 10$ -- are obtained as
described above. It is assumed that the recombining ion is in the ground
state, and recombination can take place into the ground or any of the
excited recombined (e+ion) states.

Recombination rate coefficients of
individual levels are obtained by convolving recombination cross
sections over Maxwellian electron distribution $f(v)$ at a given
temperature,
\begin{equation}
\alpha_{RC}(T) = \int_0^{\infty}{vf(v)\sigma_{RC}dv}.
\end{equation}
Contribution of these low-\en group (A) bound states to the
total $\sigma_{\rm RC}$
is obtained by summing over all low-\en group (A) states.
$\sigma_{\rm RC}$ thus obtained from $\sigma_{\rm PI}$, including
the radiatively damped autoionizing resonances (Eq. (8)), corresponds to
the total (DR+RR) unified recombination cross section.

Recombination into the high-$n$ states, $n_{\rm max} < n \leq \infty$,
(Fig.~1 of Nahar \& Pradhan 1994) are included separately. To each excited
threshold of the core, $J_i\pi_i$, belongs an infinite series of
($N$+1)-electron levels, $J_i\pi_i\nu\ell$, to which recombination
can occur. For the high $\nu$ levels DR dominates while the background
RR is negligibly small. The contributions from these levels are added
by calculating the DR collision strengths, $\Omega_{\rm DR}$, employing
the precise theory of radiation damping by Bell and Seaton (1985; Nahar
\& Pradhan 1994). The recombination cross section, $\sigma_{RC}$ in
Megabarns (Mb), is related to the collision strength, $\Omega_{\rm RC}$, as
\begin{equation}
\sigma_{RC}(i\rightarrow j)({Mb}) = \pi \Omega_{RC}(i,j)/(g_ik_i^2)
(a_o^2/1.\times 10^{-18}) ,
\end{equation}
where $k_i^2$ is the incident electron energy in Rydbergs. As $\sigma_{RC}$
diverges at zero-photoelectron energy, the total collision strength,
$\Omega$, is used in the recombination rate calculations.
Background photoionization cross sections for the high-\en
group (B) levels are computed hydrogenically, which is referred to as the
``high-\en top-up" (Nahar 1996).

\section{COMPUTATIONS}

Relativistic BPRM calculations in intermediate coupling are carried
out in the close coupling approximation using the $R$-matrix package of
codes. These codes are extensions of the Opacity Project codes
(Berrington et al.  1987) to include relativistic effects (Scott \&
Burke 1980, Scott \& Taylor 1982, Berrington et al. 1995), as
implemented under the
Iron Project (Hummer \etal 1993). The calculations span several stages of
computation using one-electron orbital wavefunctions for
the core or target ion from an updated version of the atomic
structure code SUPERSTRUCTURE, using a scaled
Thomas-Fermi-Dirac potential (Eissner et al. 1974).

The wavefunction expansion for O VI consists of 17 fine structure
levels of configurations $1s^2$, $1s2s$, $1s2p$, $1s3s$, $1s3p$, and
$1s3d$ of O VII. The levels, along with their relative energies, are given
in Table 1. The set of correlation configurations in the atomic structure
calculations and the Thomas-Fermi scaling paramenter ($\lambda_{nl}$) for
each orbital are also given in Table 1. The second term in Eq. (3), with
bound state correlation functions for O VI, includes all possible
($N$\,+\,1)-particle configurations with 0 to maximum orbital
occupancies: 2s$^2$, 2p$^2$, 3s$^2$, 3p$^2$, 3d$^2$, $4s$ and $4p$.

For O VII, the wavefunction expansion consists of 16 fine structure
levels of O VIII of $1s$, $2s$, $2p$, $3s$, $3p$, $3d$, $4s$, $4p$, $4d$
and $4f$, as given in Table 1. The bound state correlation functions
included all configurations from 0 to maximum orbital occupancies:
$1s^2$, 2s$^2$, 2p$^2$, 3s$^2$, 3p$^2$, 3d$^2$, $4s^2$ and $4p^2$,
$4d$ and $4f$. The energies in Table 1 are observed values (from the
NIST website: www.nist.gov). Although calculated energies are less than
one percent of the observed ones, the latter are used in the
computations to obtain more accurate positions of resonances.
Radial integrals for the partial wave expansion in Eq.\,3 are specified
for orbitals $0\leq\ell\leq 9$, with a R-matrix basis set
40 `continuum' functions (NRANG2) for O~VI, and 30 for O~VII.

Both the {\it partial} and the {\it total} photoionization cross setions are
obtained for all bound levels. Coupled channel calculations for
$\sigma_{\rm PI}$ include both the background and the resonance
structures (due to the doubly excited autoionizing states) in the cross
sections. Radiation damping of resonances up to $n = 10$ are included
through use of the extended codes STGF and STGBF ( Nahar and Pradhan 1994,
Zhang \etal 1999). The BPRM calculations are carried out for each total
angular momentum symmetry $J\pi$, corresponding to a set of fine
structure target levels $J_t$.  Radiation damping of resonances
within the close coupling BPRM calculations are described in Zhang \etal
(1999, and references therein). The program PBRRAD is used to extend the
{\it total} photoionization cross sections in the high energy region,
beyond the highest target threshold in the close coupling wavefunction
expansion of the ion, by a `tail' using Kramers formula $\sigma_{PI}(E) =
\sigma{PI}^o(E^o/E)^3$ where $E^o$ is the last tabulated energy above
all target thresholds.

Level-specific recombination cross sections $\sigma_{\rm RC}(i)$,
into bound levels $i \equiv$ \en(SLJ) of the recombined (e~+~ion) system,
are obtained from
{\it partial} photoionization cross sections $ \sigma_{\rm PI}(i,g)$ of
the level $i$ into the ground level $g$ of the recombining ion. These detailed
photo-recombination cross sections are calculated in the energy region
from the threshold energy up to $E(\nu = \nu_o \approx 10.0)$,
where $\nu$ is the
effective quantum number relative to the target level of the recombining
ion. The resonances up to $\nu \leq \nu_o$ are delineated with a fine energy
mesh. The electrons in this energy range generally recombine to a large
number of final (e~+~ion) levels. Recombination cross sections are
computed for all coupled symmetries and levels, and summed to obtain the
total $\sigma_{\rm RC}$.
Level specific recombination rate coefficients are obtained
using a new computer program, BPRRC (Nahar et al 2000).  The level
specific rates are obtained for energies going up to infinity.
These rates include both non-resonant and resonant contributions up to
energies $z^2/\nu_o^2$; Contributions from all autoionizing
resonances up to $\nu \leq \nu_o \approx 10$ are included.

In the higher energy region, $\nu_o < \nu < \infty$, below each
target threshold where the resonances are narrow and dense and
the background is negligible, we compute detailed and resonance
averaged DR cross sections.  The BPRM DR collision strengths are
obtained using extensions of the $R$-matrix asymptotic region code STGF
(Nahar \& Pradhan 1994, Zhang \etal 1999). It is necessary to use
a very fine energy mesh in order to delineate the resonance structures.

The program BPRRC sums up the level specific rates, which is added to
the contributions from the resonant high-n DR, from resonances with
 $\nu_o < \nu < \infty$, to obtain total recombination
rates. As an additional check on the numerical calculations,
the total recombination rate coefficients, $\alpha_R$, are also
calculated from the total recombination collision strength,
$\Omega_{RC}$, obtained from all the photoionization cross sections,
and the DR collision strengths. The agreement between the two numerical
approaches is within a few percent.

Finally, the small background (non-resonant) contribution from the
high-\en states ($10 < n \leq \infty$) to total recombination is also
included as the "top-up" part, computed in the hydrogenic approximation
(Nahar 1996). This contribution is important at low temperatures, but
negligible at high temperatures. At low temperature, recombination rate
is dominated by the RR into the infinite number of high-\en states,
as electron energies are not usually high enough for resonant excitations
and DR, and causes a rapid rise in $\alpha_R$(T) towards low temperatures.

\section{RESULTS AND DISCUSSION}

Results are presented for photoionization and recombination of O VIII
+ e $\rightarrow$ O VI, and O VIII + e $\rightarrow$ O VII.  Odd and
even parity energy levels \en(SLJ) of total angular momentum symmetries
$1/2 \leq J \leq 17/2$ are considered for O~VI, and
$0 \leq J \leq 4$ for O~VII, all bound levels with \en $\leq 10$.

Photoionization cross sections $\sigma_{PI}$ include both the {\it total}
cross section, leaving the core (residual) ion in the ground and
excited levels, as well as the {\it partial} cross sections, leaving
the core ion in the ground level only. The total cross sections are
needed in astrophysical applications, such as in ionization balance
calculations while parital cross sections are needed for applications
such as for recombination rate coefficients. (Both the total and the
partial cross sections are available electronically.)

Level-specific and total recombination rate coefficients using the
BPRM unified treatment, $\alpha_i$ (\en SLJ, \en $\leq$ 10) and
$\alpha_{RC}(T)$ respectively, are presented for O~VI and O~VII. In the
following subsections we describe salient features pertaining to
$\alpha_{RC}(T)$ in numerical form and in figures. Total $\alpha_{RC}(T)$
are computed two different ways to enable numerical checks: (i) from
the sum of the level-specific rate coefficients and the high-n DR
contribution, and (ii) from total collision strengths calculated from
photoionization cross sections directly, and the DR contribution. The
differences between the two are typically within a few percent, thus
providing a numerical and self-consistency check particularly on the
resolution of resonances.  We also present total recombination rate
coefficients for the hydrogenic ion O~VIII for completeness.

\subsection{O VI}

\subsubsection{Photoionization}

Figs. 1(a) and (b) show the ground state photoionization cross
section for O VI ($1s^2 \ 2s \ ^2S_{1/2}$). The top panel (a) presents
the total photoionization cross section summed over the various
target thresholds for ionization, and the bottom panel (b) presents
the partial cross sections of the ground level into the ground
$1s^2(^1S)$ level of residual ion O VII. The resonances at high
energies belong to Rydberg series converging $n=2,3$ levels. These are
the well known KLL, KLn (\en $>$ 2) as, for example, discussed by
Pradhan (2000) and Nahar \etal (2001). Since the first excited levels
of n=2 thresholds of the core ion O~VII lie at high energies, the
cross sections decrease monotonically over a large energy range before
the Rydberg series of resonances appears. The total and the
partial cross sections are identical below the first excited level of
the residual ion beyond which total $\sigma_{PI}$ increases due to
added contributions from excited channels, as shown in the figures. A
distinct difference between the total and partial cross sections
Fig. 1 comes from the contribution
of channels with excited n=2 thresholds. The K-shell ionization
jump at the n = 2 target levels in total $\sigma_{PI}$ is due to
inner-shell photoionization:
$$ h\nu + O~VI (1s^2 \ 2s) \longrightarrow e \ + O~VII (1s2s ,1s2p) \ .$$
In X-ray photoionization models inner-shell edges play an important
role in overall ionization rates.

Fig. 2 presents partial photoionization cross sections of the Rydberg
series of levels $1s^2ns$, 2 $\leq n \leq$ 10. As noted in earlier
works, photoionization cross section of Rydberg levels exhibit
wide resonances known as photoexcitation-of-core (PEC) resonances at
energies associated with dipole transitions in the core ion. We see
such PEC resonances in in cross sections of all bound levels of O~VI, as
in Fig. 2 at photon energies 41.788, 42.184, 48.804 and 48.922 Ry due
to core excitations to levels $1s2p(^3P^o_1)$, $1s2p(^1P^o_1)$,
$1s3p(^3P^o_1)$, and $1s3p(^1P^o_1)$ of O~VII. At these energies the
core ion goes through an allowed transition, while the outer electron
remains a `spectator' in a doubly-excited resonance state, followed
by autoionization of the O~VI resonance into the ground level of O~VII.
The effect gets more distinct for cross sections of higher excited
levels. For low charge ions, PEC resonances are usualy wider. These
resonances in O~VI depict the behavior of cross sections that determine
corresponding features in level-specific recombination rates.

\subsubsection{Recombination cross sections and rate coefficients}

The results consist of the total and 98 level-specific ($n \leq$ 10)
recombination rate coefficients $\alpha_R(T)$ and $\alpha_i(T)$
respectively. Total recombination rates are given in Table 2. The main
features are illustrated and compared with previously available data for
RR and DR rates in Fig. 3. The solid curve is the BPRM total unified
$\alpha_R(T)$ and shows typical features. The high values at very low
temperature are due to the dominance of RR into an infinity of high-\en
levels. $\alpha_R(T)$ decreases with increasing T until high
temperatures where it rises due to the dominance of DR, followed by a
monotonous decrease. The dotted curve in Fig. 3 is the total rate in LS
couplig (Nahar 1999). The difference between the LS coupling
$\alpha_R(T)$ and the present BPRM rate is due to relativistic effects
{\it and} radiation damping of low-\en resonances.
Both effects can be important for highly charged ions.
The dashed curve is the RR
rate by Pequignot et al (1991) obtained by smoothing of OP data over
resonances, and
hydrogenic approximation. The agreement is good in the low region since
non-resonant cross sections dominate the low energy region. The
dot-dashed curve is the DR rate by Badnell et al. (1990) using the
isolated resonance
approximation. The sum of previous RR and DR rates agrees with the
present unified rates to about 10-20\%, as expected for highly charged
ions.

The 98 level-specific recombination rate coefficients $\alpha_i(T)$ for
O VI are for $i \equiv n(SLJ)$, n $\leq$ 10 and $\ell \leq 9$, and
associated $J\pi$ levels. Fig. 4 presents $\alpha_i(T)$ into eight
\en = 2 and 3 levels with \en(SLJ): $2s~^2S_0$, $2p~^2P^o_{1/2,3/2}$,
$3s~^2S_0$, $3p~^2P^o_{1/2,3/2}$, and $3d~^2D_{3/2,5/2}$. These rates
are relatively smooth except for a small and diffuse DR `bump'.
The levels correspond to the prominent 1032 and 1038 \ang UV doublet
lines respectively in O~VI spectra arising due to strong dipole
transitions $^2P^o_{3/2,1/2} \longrightarrow ^2S_{1/2}$.

Total photorecombination cross sections $\sigma_{RC}$ show features
similar to photoionization cross sections $\sigma_{PI}$ ---
smooth decay with energy,
before the resonance complexes begin to emerge at high energies. However,
($\sigma_{RC}$ are more complicated than $\sigma_{PI}$ since all levels
contribute to the total $\sigma_{RC}$, shown in Fig 5(a) for O VI.
The resonance complexes are marked as KLL, KLM, KLN etc. For heavier
high-Z elements than oxygen (Z $>$ 10) these manifest themselves as
di-electronic satellite lines observed in tokamaks,
Electron-Beam-Ion-Traps (EBIT), ion storage rings and astrophysical
sources. The KLL complexes have been well studied in previous works
(e.g. Gabriel 1972, Bely-Dubau \etal 1982, Pradhan and Zhang 1997, Zhang et al.
1999), Beiersdorfer \etal 1992, Oelgoetz and Pradhan 2001), for various ions.

\subsection{O VII}

\subsubsection{Photoionization}

Total and partial photoionization cross sections $\sigma_{PI}$
for 116 levels \en(SLJ) ( $n \leq$ 10) are computed (the partial cross
sections are for ionization into the ground level 1s($^2S_{1/2}$) of
O~VIII).  Illustrative results are presented in Fig. 6.  Fig. 6(a) shows
the photoionization cross section of the ground level $(1s^2 \ ^1S_0)$
of O VII. Similar to O VI, the Rydberg series of resonances, KL and
K\en ($n > 2$), begins at fairly high energies owing to the high
$n=2$ excitation thresholds of O~VIII. However, the ground level
$\sigma_{PI}$ of O~VII does not show a significant K-shell jump at $n=2$
threshold, as seen in O~VI. Nonetheless the feature is prominent in the
excited \en = 2 level cross sections as seen in Figs. 6(b)-(e). These
excited levels are responsible for the formation of the important X-ray
diagnostic lines, w, x, y, z, from the allowed and forbidden transitions
$1s^2 \ (^1S_0) \longleftarrow 1s2p (^1P^o_1), 1s2p (^3P^o_2), 1s2p
(^3P^o_1), 1s2s (^3S_1)$, respectively. The K-shell
ionization jump at the n = 2 target levels
$$ h\nu + O~VII (1s2s, 1s2p) \longrightarrow e \ + O~VIII (2s \ ,2p) \ . $$
can be seen clearly in the photoionization cross sections.

\subsubsection{Recombination cross sections and rate coefficients}

Total recombination rate coefficients for O VII over a wide temperature
range are presented in Table 3. In Fig. 7 the solid curve is
the present total recombination rate coefficients, including
relativistic effects in BPRM approximation and radiation damping of
low-n ($n \leq 10$) resonances. The solid curve has the same general shape
typical of unified recombination rates - the high rate at low temperature
deceases until the recombination is dominated by DR forming a hump at
high temperature. The dotted curve in the figure is the earlier
total unified rate (Nahar 1999) where a part of the rates, the
resonances $2\ell n\ell$ converging on to the $n = 2$ thresholds of O~VIII,
were also computed using the BPRM method including radiation damping
(Zhang et al. 1999); the remainder was in LS coupling. Both results
show very close agreement since the main $n = 2$ resonant contributions
are the same. However, present rates also include $n = 3,4$
resonant contributions $3\ell n\ell, 4\ell n \ell$.
The dashed curve is the  RR rate obtained by Verner and Ferland (1996),
which agrees well with the present one in the region where DR is small.
The dot-dashed curve is the DR rate calculated by Savin (1999) from
experimentally measured recombination cross sections by Wolf et al.
(1991). As is discernible from Fig. 7, the sum of the previous RR and
the experimental DR rates (RR+DR) is significantly lower than the
present unified rate for O~VII. At the temperature of peak DR
contribution Log T = 6.6, the unified rate is about 20\% higher the
RR+DR value.
 This is most likely because the experimental DR measurements cover
a smaller energy range and do not include all contributions.

Level-specific recombination rate coefficients are obtained for 116
levels \en(SLJ) with 0  $ \leq J \leq$ 4 and \en $\leq$ 10.
These rates for fine structure levels were not obtained in our earlier
work (Nahar 1999). Fig. 8 presents level specific rates for the $n = 2$
levels corresponding to the X-ray w, x, y, and z lines of O VII. The
rates show a relatively smooth decay with temperature except for the
high temperature DR bump for $1s2p(^1P^o_1)$ and $1s2s(^3S_1)$ levels.

The total recombination cross sections $\sigma_{RC}$, summed over
recombination cross sections of individual fine structure levels up to
n = 10 of O VII and high-n contributions are presented in Fig. 5(b).
The resonance complexes LL, LM, LN, MM etc. of n = 2 and 3 thresholds
of the core are specified
in the figure. The $n = 4$ resonances are too low for any significant
contributions. These complexes are seen at high energies near the high
lying O~VIII thresholds.

As mentioned above, as a numerical check on the calculations we verify
that the sum of the level-specific rate coefficients
and the DR contribution agrees within a few percent with the total
recombination rate coefficient obtained from total collision strengths
obtained from $\sigma_{RC}$ in Fig. 5 for both O VI and O VII.

\subsubsection{Ionization fractions in coronal equilibrium}

Using the present total unified recombination rates, $\alpha_R(T)$,
for O~VI and O~VII we
calculate ionization fractions of oxygen ions in coronal
(collisional) equilibrium,
\begin{equation}
N(z-1,g)S(z-1,g) = N(z,g)\alpha(z,g),  1\leq z\leq z_{max},
\end{equation}
where $S$ is the rate coefficient for electron impact ionization and
total element density $N_T = \sum_{z=0}^{z_{max}}{N(z,g)}$. The
ionization fractions in $-log[N(z)/N_T]$ are given in Table 3. However,
present fractions show very little difference from earlier values (Nahar
1999). Most of the differencs are in the third significant digit and do
not show up in Table 3.

\section{CONCLUSION}

Extensive results from relativistic calculations for total and
level-specific photoionization and recombination cross sections and rates
are presented for O~VI and O~VII. These are of general interest in UV and
X-ray spectroscopy of laboratory and astrophysical sources, especially
the formation due to recombination of important lines such as
the $\lambda\lambda$ 1032,1038
fine structure doublet in O~VI, and the `triplet' X-ray features in
O~VII.

A discussion of some of the important atomic effects
such as resolution and radiation damping of resonances, interference
between non-resonant (RR) and resonant (DR) recombination, comparison
with experimental data and uncertainties, and general features of the
unified (e~+~ion) recombination rates, has also been given in the first
paper on the new BPRM calculations -- paper IV in the present series
(Nahar \etal 2000).

Di-electronic satellite rates for the KLL, KLM, etc, complexes of
several ions have earlier been shown to be in very good agreement with
experiments and other theoretical calculations (Pradhan and Zhang 1997;
Zhang \etal 1999), to about 10-20\%; it is therefore expected that the
present rates should be definitive, with similar uncertainties.

The present level-specific data can be used to construct recombination-cascade
matrices for O~VI and O~VII,  to obtain effective recombination rates
into specific fine structure levels \en(SLJ)
with $n \leq 10$ and $\ell \leq n-1 $ (e.g. Pradhan 1985).
The present data is more than sufficient
for extrapolation to high-n,$\ell$ necessary to account for
all cascade contributions.

The available data includes:
(A) Photoionization cross sections, both total and partial, for bound fine
structure levels of O~VI and O~VII up to the $n = 10$ levels.
(B) Total unified recombination rates for O~VI and O VII, and
level-specific recombination rate coefficients for levels up to $n = 10$.
Further calculations for other He-like and Li-like ions are in progress.
All photoionization and recombination data are available electronically
at: nahar@astronomy.ohio-state.edu. The total
recombination rate coefficients are also available from the Ohio State
Atomic Astrophysics website at: www.astronomy.ohio-state.edu/$\sim$pradhan.

%

\acknowledgments

This work was supported partially by NSF and NASA. The
computational work was carried out on the Cray SV1 at the Ohio Supercomputer
Center in Columbus, Ohio.

\clearpage

\begin{table}
\caption{Target levels in the eigenfunction expansions of O VII and O
VIII.
}
\scriptsize
\begin{tabular}{rllrll}
\hline
&\multicolumn{2}{c}{O VI} & \multicolumn{3}{c}{O VIII} \\
& level & $E_t(Ry)$ & & level & $E_t(Ry)$ \\
\hline
1&1s$^2(^1{\rm S}_0)$     & 0.0    &
1&1s$(^2{\rm S}_{1/2})$   &   0.00 \\
2&1s2s$(^3{\rm S}_1)$     &  41.232    &
2&2p$(^2{\rm P}^o_{1/2})$   & 48.0308997\\
3&1s2p$(^3{\rm P^o}_0)$     & 41.787     &
3&2s$(^2{\rm S}_{1/2})$ & 48.0315698\\
4&1s2p$(^3{\rm P}^o_1)$   &   41.788   &
4&2p$(^2{\rm P}^o_{3/2})$ &48.0445917\\
5&1s2p$(^3{\rm P}^o_2)$   &  41.793    &
5&3p$(^2{\rm P}^o_{1/2})$ &56.9304970 \\
6&1s2s$(^1{\rm S}_0)$   &  41.812    &
6&3s$(^2{\rm S}_{1/2})$   &56.9306973 \\
7&1s2p$(^1{\rm P}^o_1)$   &  42.184    &
7&3d$(^2{\rm D}_{1/2})$ & 56.9345473 \\
8&1s3s$(^3{\rm S}_1)$     &  48.651    &
8&3p$(^2{\rm P}^o_{3/2})$ &56.9345541\\
9&1s3p$(^3{\rm P}_0)$     &  48.804    &
9&3d$(^2{\rm D}_{5/2})$   &56.9358976\\
10&1s3p$(^3{\rm P}^o_1)$   &  48.804    &
10&4p$(^2{\rm P}^o_{1/2})$ &60.0447993 \\
11&1s3p$(^3{\rm P}^o_2)$   &  48.804  &
11&4s$(^2{\rm S}_{1/2})$   &60.0448841\\
12&1s3s$(^1{\rm S}_0)$   &  48.811    &
12&4d$(^2{\rm D}_{3/2})$   &60.0465078\\
13&1s3d$(^3{\rm D}_3)$   &  48.884    &
13&4p$(^2{\rm P}^o_{3/2})$ &60.0465108 \\
14&1s3d$(^3{\rm D}_2)$   &  48.884    &
14&4f$(^2{\rm P}^o_{5/2})$ &60.0470765 \\
15&1s3d$(^3{\rm D}_1)$   &  48.884    &
15&4d$(^2{\rm D}_{5/2})$ &60.0470775\\
16&1s3d$(^1{\rm D}_1)$   &  48.894    &
16&4f$(^2{\rm P}^o_{7/2})$&60.0473612 \\
17&1s3p$(^1{\rm P}^o_1)$   & 48.922   &
 & \\
\hline
\multicolumn{6}{l}{O VII: Correlations -  $2s^2$, $2p^2$, $3s^2$,
$3p^2$, $3d^2$, $2s2p$, $2s3s$,}\\
\multicolumn{6}{l}{$2s3p$,$2s3d$, $2s4s$, $2s4p$, $2p3s$,
$2p3p$, $2p3d$, $2p4s$, $2p4p$,} \\
\multicolumn{6}{l}{$\lambda_{nl}$ - 0.991(1s),0.991(2s),0.776(2p),1.16883(3s),0.91077(3p),} \\
\multicolumn{6}{l}{1.00746(3d),-1.59699(4s),-1.61237(4p) --- see text)} \\
\multicolumn{6}{l}{O VIII: $\lambda_{nl}$ - 1.0, for 1s to 4f } \\
\end{tabular}
\end{table}

\pagebreak

\begin{table}
\caption{Total recombination rate coefficients $\alpha_R(T)$ for O~VI,
O~VII and O~VIII. }
\scriptsize
\begin{tabular}{crrrcrrr}
\hline
$log_{10}T$ & \multicolumn{3}{c}{$\alpha_R(cm^3s^{-1})$} &
$log_{10}T$ & \multicolumn{3}{c}{$\alpha_R(cm^3s^{-1})$}\\
(K) & \multicolumn{1}{c}{O VI} & \multicolumn{1}{c}{O VII} &
\multicolumn{1}{c}{O VIII} & (K) & \multicolumn{1}{c}{O VI} &
\multicolumn{1}{c}{O VII} & \multicolumn{1}{c}{O VIII} \\
\hline
  1.0&    1.42E-09 & 1.99E-09 & 2.90E-09 &
  5.1&    3.72E-12 & 6.16E-12 & 1.03E-11 \\
  1.1&    1.25E-09 & 1.76E-09 & 2.56E-09 &
  5.2&    3.13E-12 & 5.28E-12 & 8.78E-12 \\
  1.2&    1.10E-09 & 1.55E-09 & 2.26E-09 &
  5.3&    2.62E-12 & 4.51E-12 & 7.51E-12 \\
  1.3&    9.68E-10 & 1.36E-09 & 2.00E-09 &
  5.4&    2.19E-12 & 3.85E-12 & 6.42E-12 \\
  1.4&    8.50E-10 & 1.20E-09 & 1.76E-09 &
  5.5&    1.83E-12 & 3.28E-12 & 5.48E-12 \\
  1.5&    7.46E-10 & 1.05E-09 & 1.55E-09 &
  5.6&    1.52E-12 & 2.79E-12 & 4.67E-12   \\
  1.6&    6.53E-10 & 9.23E-10 & 1.36E-09 &
  5.7&    1.26E-12 & 2.37E-12 & 3.97E-12 \\
  1.7&    5.72E-10 & 8.10E-10 & 1.20E-09 &
  5.8&    1.05E-12 & 2.01E-12 & 3.37E-12   \\
  1.8&    5.01E-10 & 7.10E-10 & 1.05E-09 &
  5.9&    8.99E-13 & 1.72E-12 & 2.85E-12 \\
  1.9&    4.38E-10 & 6.21E-10 & 9.26E-10 &
  6.0&    8.42E-13 & 1.52E-12 & 2.40E-12   \\
  2.0&    3.83E-10 & 5.44E-10 & 8.12E-10 &
  6.1&    9.14E-13 & 1.44E-12 & 2.03E-12 \\
  2.1&    3.34E-10 & 4.75E-10 & 7.12E-10 &
  6.2&    1.11E-12 & 1.52E-12 & 1.71E-12   \\
  2.2&    2.91E-10 & 4.15E-10 & 6.23E-10 &
  6.3&    1.39E-12 & 1.73E-12 & 1.43E-12 \\
  2.3&    2.54E-10 & 3.62E-10 & 5.46E-10 &
  6.4&    1.65E-12 & 1.99E-12 & 1.20E-12   \\
  2.4&    2.21E-10 & 3.16E-10 & 4.78E-10 &
  6.5&    1.82E-12 & 2.21E-12 & 9.96E-13 \\
  2.5&    1.93E-10 & 2.76E-10 & 4.18E-10 &
  6.6&    1.86E-12 & 2.31E-12 & 8.27E-13   \\
  2.6&    1.67E-10 & 2.40E-10 & 3.65E-10 &
  6.7&    1.78E-12 & 2.26E-12 & 6.84E-13 \\
  2.7&    1.46E-10 & 2.09E-10 & 3.19E-10 &
  6.8&    1.60E-12 & 2.10E-12 & 5.63E-13   \\
  2.8&    1.27E-10 & 1.82E-10 & 2.78E-10 &
  6.9&    1.38E-12 & 1.85E-12 & 4.61E-13 \\
  2.9&    1.10E-10 & 1.58E-10 & 2.43E-10 &
  7.0&    1.14E-12 & 1.56E-12 & 3.75E-13   \\
  3.0&    9.52E-11 & 1.37E-10 & 2.12E-10 &
  7.1&    9.19E-13 & 1.28E-12 & 3.05E-13 \\
  3.1&    8.25E-11 & 1.19E-10 & 1.85E-10 &
  7.2&    7.20E-13 & 1.02E-12 & 2.45E-13   \\
  3.2&    7.15E-11 & 1.03E-10 & 1.61E-10 &
  7.3&    5.54E-13 & 7.96E-13 & 1.96E-13 \\
  3.3&    6.19E-11 & 8.97E-11 & 1.40E-10 &
  7.4&    4.19E-13 & 6.10E-13 & 1.57E-13   \\
  3.4&    5.35E-11 & 7.77E-11 & 1.22E-10 &
  7.5&    3.13E-13 & 4.61E-13 & 1.24E-13 \\
  3.5&    4.62E-11 & 6.73E-11 & 1.06E-10 &
  7.6&    2.32E-13 & 3.45E-13 & 9.76E-14   \\
  3.6&    3.99E-11 & 5.82E-11 & 9.22E-11 &
  7.7&    1.70E-13 & 2.56E-13 & 7.63E-14 \\
  3.7&    3.43E-11 & 5.04E-11 & 8.01E-11 &
  7.8&    1.24E-13 & 1.89E-13 & 5.93E-14   \\
  3.8&    2.96E-11 & 4.35E-11 & 6.95E-11 &
  7.9&    9.00E-14 & 1.38E-13 & 4.59E-14 \\
  3.9&    2.54E-11 & 3.76E-11 & 6.03E-11 &
  8.0&    6.50E-14 & 1.01E-13 & 3.51E-14   \\
  4.0&    2.18E-11 & 3.25E-11 & 5.22E-11 &
  8.1&    4.69E-14 & 7.35E-14 & 2.70E-14 \\
  4.1&    1.87E-11 & 2.80E-11 & 4.53E-11 &
  8.2&    3.37E-14 & 5.33E-14 & 2.05E-14   \\
  4.2&    1.61E-11 & 2.41E-11 & 3.92E-11 &
  8.3&    2.42E-14 & 3.86E-14 & 1.55E-14 \\
  4.3&    1.37E-11 & 2.08E-11 & 3.39E-11 &
  8.4&    1.73E-14 & 2.79E-14 & 1.16E-14   \\
  4.4&    1.17E-11 & 1.79E-11 & 2.93E-11 &
  8.5&    1.24E-14 & 2.02E-14 & 8.72E-15 \\
  4.5&    1.00E-11 & 1.54E-11 & 2.53E-11 &
  8.6&    8.87E-15 & 1.45E-14 & 6.51E-15   \\
  4.6&    8.54E-12 & 1.32E-11 & 2.18E-11 &
  8.7&    6.33E-15 & 1.05E-14 & 4.83E-15 \\
  4.7&    7.26E-12 & 1.14E-11 & 1.88E-11 &
  8.8&    4.52E-15 & 7.55E-15 & 3.58E-15   \\
  4.8&    6.16E-12 & 9.77E-12 & 1.62E-11 &
  8.9&    3.23E-15 & 5.44E-15 & 2.64E-15 \\
  4.9&    5.22E-12 & 8.38E-12 & 1.39E-11 &
  9.0&    2.30E-15 & 3.91E-15 & 1.93E-15   \\
  5.0&    4.41E-12 & 7.20E-12 & 1.19E-11 &
  & & & \\
\hline
\end{tabular}
\end{table}

\begin{table}
\caption{Ionization fractions, $-log_{10}{N(z)\over N_T}$, of oxygen ions
in coronal equilibrium. }
\scriptsize
\begin{tabular}{cccccccccc}
\hline
$log_{10}T$ & \multicolumn{9}{c}{$-log_{10}{N(z)/N_T}$} \\
 & O I & O II & O III & O IV & O V & O VI & O VII & O VIII & O IX \\
\hline
 4.0 & 0.001 & 2.5   & ***** & ***** & ***** & ***** & ***** & ***** & *****\\
  4.1 & 0.0433& 1.02  & ***** & ***** & ***** & ***** & ***** & ***** & *****\\
  4.2 & 0.432 & 0.2   & 7.91  & ***** & ***** & ***** & ***** & ***** & *****\\
  4.3 & 1.19  & 0.029 & 5.37  & ***** & ***** & ***** & ***** & ***** & *****\\
  4.4 & 1.77  & 0.008 & 3.49  & ***** & ***** & ***** & ***** & ***** & *****\\
  4.5 & 2.12  & 0.007 & 2.10  & 8.15  & ***** & ***** & ***** & ***** & *****\\
  4.6 & 2.38  & 0.037 & 1.11  & 5.38  & ***** & ***** & ***** & ***** & *****\\
  4.7 & 2.72  & 0.176 & 0.481 & 3.39  & 9.73  & ***** & ***** & ***** & *****\\
  4.8 & 3.21  & 0.483 & 0.179 & 2.05  & 6.89  & ***** & ***** & ***** & *****\\
  4.9 & 3.82  & 0.901 & 0.096 & 1.14  & 4.76  & ***** & ***** & ***** & *****\\
  5.0 & 4.52  & 1.40  & 0.174 & 0.539 & 3.17  & 7.54  & ***** & ***** & *****\\
  5.1 & 5.33  & 2.01  & 0.417 & 0.23 &  2.02  & 5.13  & 8.58  & ***** & *****\\
  5.2 & 6.22  & 2.71  & 0.778 & 0.114 & 1.21  & 3.28  & 5.48  & ***** & *****\\
  5.3 & 7.16  & 3.45  & 1.22  & 0.155 & 0.646 & 1.87  & 3.04  & ***** & *****\\
  5.4 & 8.23  & 4.32  & 1.82  & 0.41  & 0.387 & 0.895 & 1.23  & ***** & *****\\
  5.5 & 9.78  & 5.67  & 2.92  & 1.20  & 0.724 & 0.634 & 0.289 & ***** & *****\\
  5.6 & ***** & 7.43  & 4.43  & 2.43  & 1.57  & 0.975 & 0.064 & 8.40  & *****\\
  5.7 & ***** & 9.11  & 5.89  & 3.64  & 2.43  & 1.40  & 0.019 & 6.32  & *****\\
  5.8 & ***** & ***** & 7.19  & 4.71  & 3.18  & 1.78  & 0.007 & 4.66  & *****\\
  5.9 & ***** & ***** & 8.33  & 5.63  & 3.82  & 2.10  & 0.004 & 3.31 &7.86 \\
  6.0 & ***** & ***** & 9.31  & 6.41  & 4.34  & 2.34  & 0.005 & 2.23 &5.52 \\
  6.1 & ***** & ***** & ***** & 7.05  & 4.75  & 2.49  & 0.020 & 1.39 &3.66 \\
  6.2 & ***** & ***** & ***** & 7.61  & 5.09  & 2.60  & 0.080 & 0.795& 2.24 \\
  6.3 & ***** & ***** & ***** & 8.19  & 5.47  & 2.76  & 0.231 & 0.455& 1.22 \\
  6.4 & ***** & ***** & ***** & 8.89  & 5.98  & 3.08  & 0.527 & 0.368& 0.56 \\
  6.5 & ***** & ***** & ***** & 9.74  & 6.65  & 3.56  & 0.984 & 0.51 & 0.231 \\
  6.6 & ***** & ***** & ***** & ***** & 7.41  & 4.15  & 1.52  & 0.77 & 0.096 \\
  6.7 & ***** & ***** & ***** & ***** & 8.16  & 4.75  & 2.05  & 1.06 &0.044 \\
  6.8 & ***** & ***** & ***** & ***** & 8.90  & 5.33  & 2.55  & 1.33 &0.022 \\
  6.9 & ***** & ***** & ***** & ***** & 9.59  & 5.88  & 3.00  & 1.57 &0.012 \\
  7.0 & ***** & ***** & ***** & ***** & ***** & 6.41  & 3.42  & 1.79 &0.007 \\
  7.1 & ***** & ***** & ***** & ***** & ***** & 6.91  & 3.81  & 1.98 &0.005 \\
  7.2 & ***** & ***** & ***** & ***** & ***** & 7.39  & 4.17  & 2.16 & 0.003 \\
  7.3 & ***** & ***** & ***** & ***** & ***** & 7.85  & 4.50  & 2.32 & 0.002 \\
  7.4 & ***** & ***** & ***** & ***** & ***** & 8.28  & 4.82  & 2.47 &0.001 \\
  7.5 & ***** & ***** & ***** & ***** & ***** & 8.70  & 5.12  & 2.61 &0.001 \\
  & & & &&&&&&\\
\hline
\end{tabular}
\end{table}


\clearpage

%
%

\def\amp{{Adv. At. Molec. Phys.}\ }
\def\apj{{ Astrophys. J.}\ }
\def\apjs{{Astrophys. J. Suppl. Ser.}\ }
\def\apjl{{Astrophys. J. (Letters)}\ }
\def\aj{{Astron. J.}\ }
\def\aa{{Astron. Astrophys.}\ }
\def\aasup{{Astron. Astrophys. Suppl.}\ }
\def\adndt{{At. Data Nucl. Data Tables}\ }
\def\cpc{{Comput. Phys. Commun.}\ }
\def\jqsrt{{J. Quant. Spect. Radiat. Transfer}\ }
\def\jpb{{Journal Of Physics B}\ }
\def\pasp{{Pub. Astron. Soc. Pacific}\ }
\def\mn{{Mon. Not. R. astr. Soc.}\ }
\def\pra{{Physical Review A}\ }
\def\prl{{Physical Review Letters}\ }
\def\zpds{{Z. Phys. D Suppl.}\ }
\def\adndt{Atomic Data And Nuclear Data Tables}

%

\begin{figure}
\vspace*{-2.0cm}
\centering
\psfig{figure=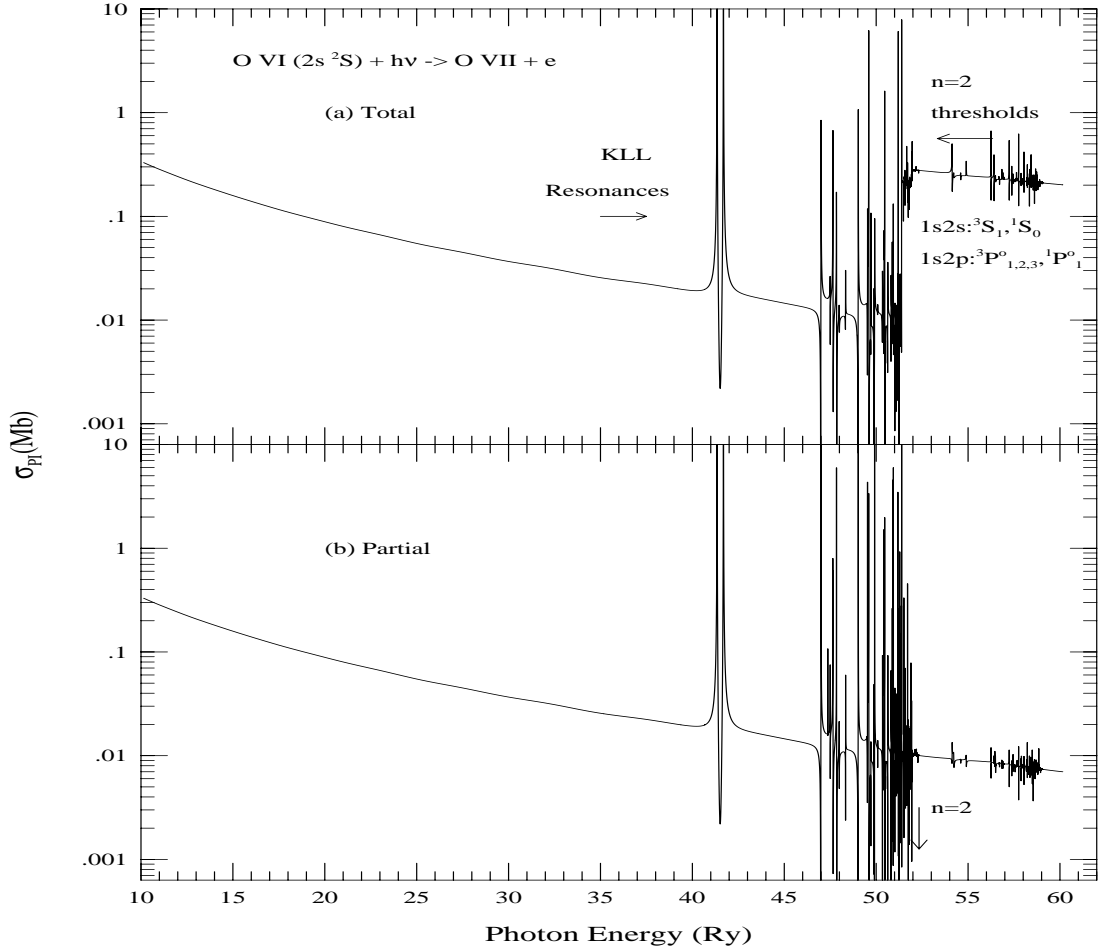,height=15.0cm,width=18.0cm}
\vspace*{-1cm}
\caption{Photoionization cross sections of the ground level
$1s^2 \ 2s \ (^2S_{1/2})$ of O~VI: (a) Total cross section; the large jump
around 52 Ry corresponds to the n=2 K-shell ionization edge. (b) Partial
cross section into the ground level $1s^2 \ (^1S_0)$ of O~VII; note that
the jump is no longer present and the cross section is continuous
across the n = 2 thresholds of O~VII.}
\end{figure}

\begin{figure}
\vspace*{-2.0cm}
\centering
\psfig{figure=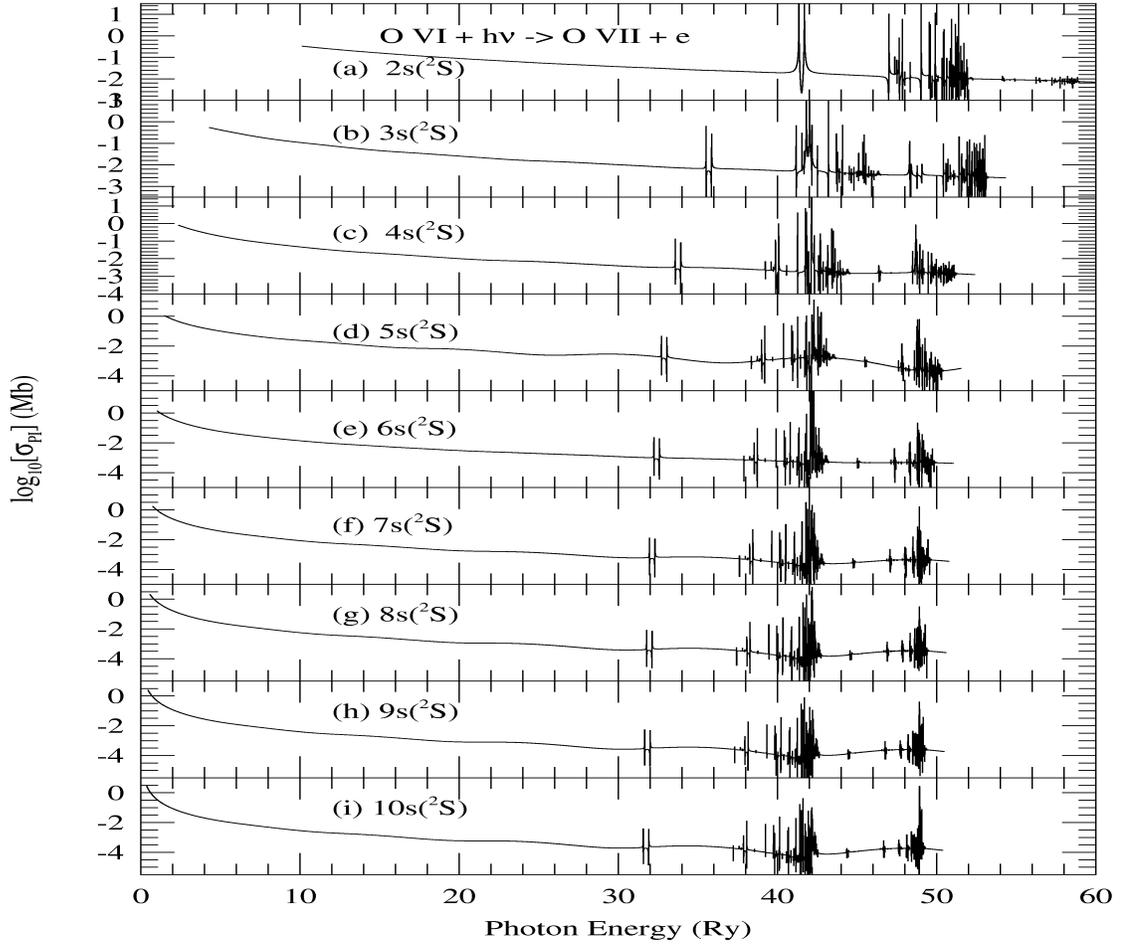,height=15.0cm,width=18.0cm}
\vspace*{-1cm}
\caption{Partial photoionization cross sections of the Rydberg series of
levels $1s^2ns(^2S_{1/2})$ of O~VI into the ground state state
$1s^2(^1S_0)$ of O~VII. {\it photoexcitation-of-core} (PEC)
resonances are seen at about 42 and 49 Ry.}
\end{figure}

\begin{figure}
\centering
\psfig{figure=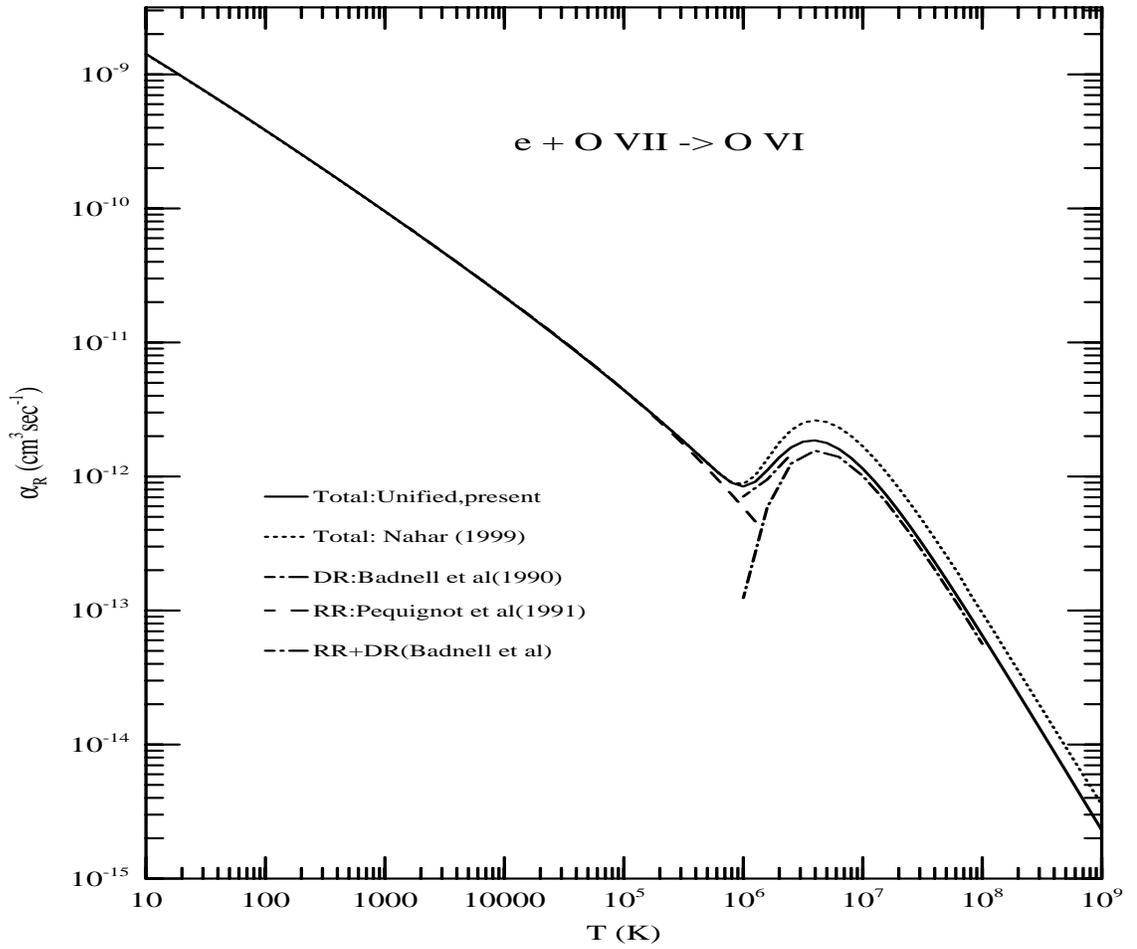,height=15.0cm,width=18.0cm}
\caption{ Total unified recombination rate coefficients for O~VI: BPRM
with fine structure (solid curve); Total in LS coupling (Nahar 1999,
dotted): RR rates by Pequignot et al (1991, dash), DR rates by Badnell
et al. (1990, dot-dash).}
\end{figure}

\begin{figure}
\psfig{figure=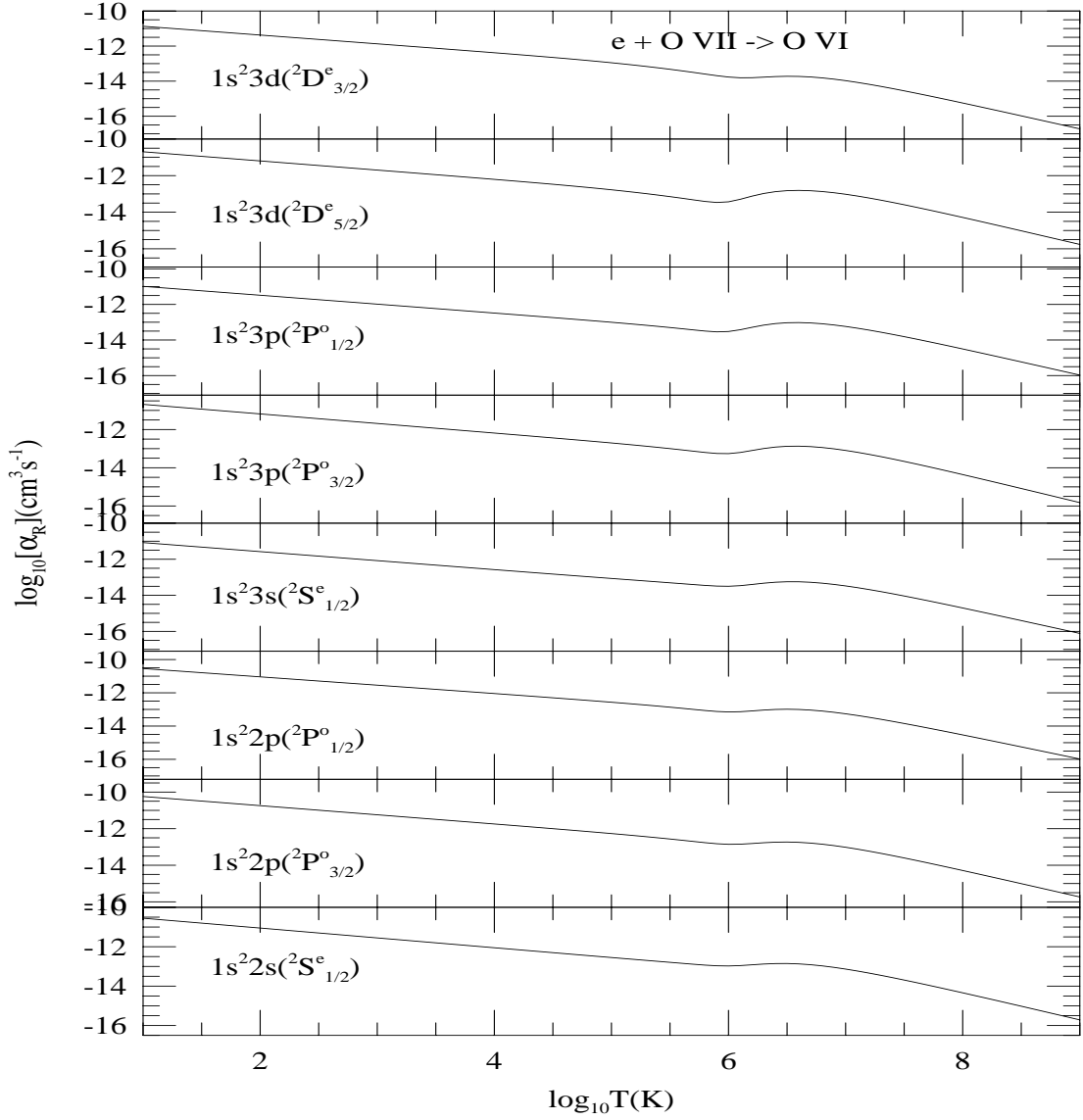,height=18.0cm,width=18.0cm}
\caption{ Level-specific recombination rate coefficients for O VI
recombining to ground and excited n=2, 3 levels. The strong dipole
transitions $^2P^o_{3/2,1/2} \longrightarrow ^2S_{1/2}$ are responsible
for the prominent 1032 and 1038 \ang UV doublet lines respectively in
O~VI spectra.}
\end{figure}

\begin{figure}
\centering
\psfig{figure=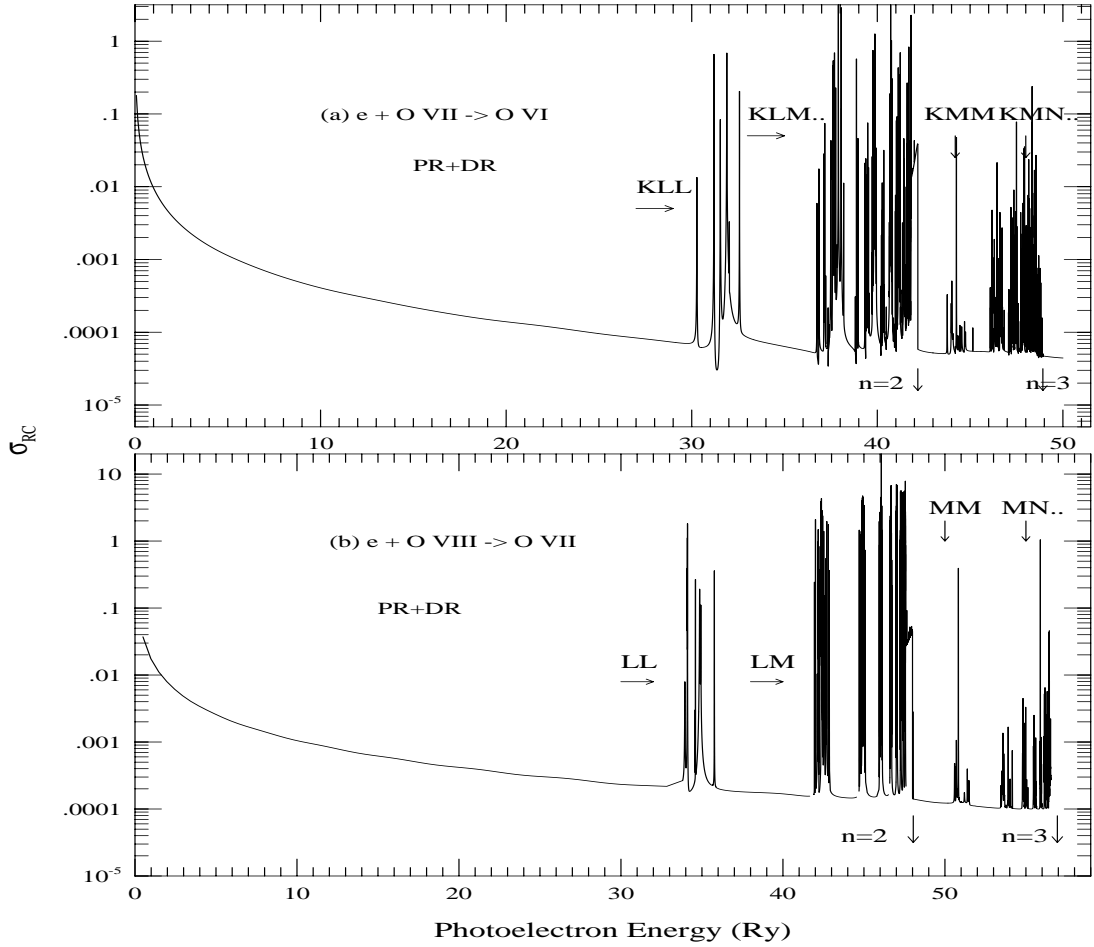,height=15.0cm,width=18.0cm}
\caption{Total unified (e + ion) photo-recombination cross sections,
$\sigma_{RC}$, of (a) O VI and (b) O VII. Note that the $\sigma_{RC}$
exhibit considerably more resonance structures than the corresponding
ground level $\sigma_{PI}$ in Figs. 1 and 6, since the former are
summed over the ground {\it and} many excited recombined levels.}
\end{figure}

\begin{figure}
\centering
\psfig{figure=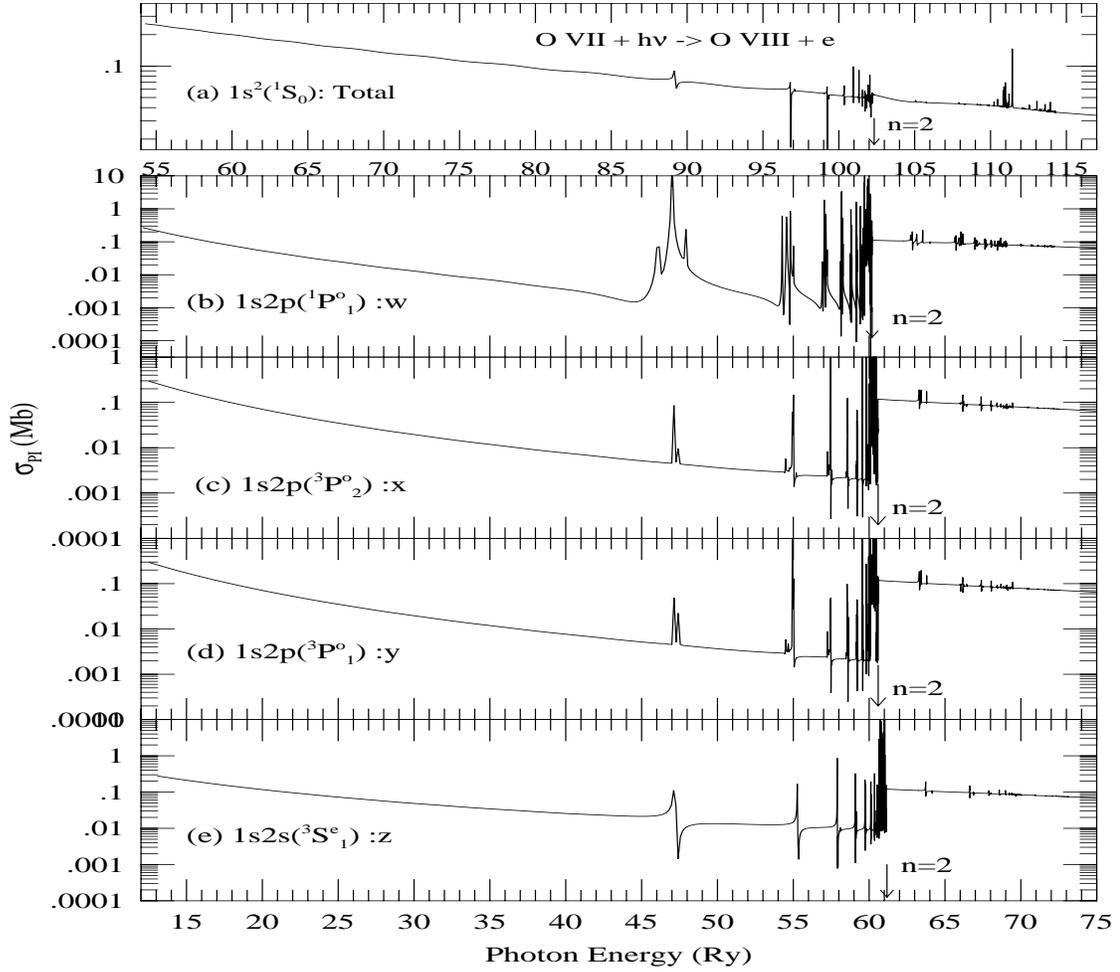,height=15.0cm,width=18.0cm}
\caption{Level-specific photoionization cross sections of (a) the ground level
$1s^2 \ (^1S_0)$, and (b) - (e) excited 1s2s and 1s2p levels of O~VII
responsible for the prominent X-ray lines: resonance (w),
intercombination (x,y), and forbidden(z).}
\end{figure}

\begin{figure}
\centering
\psfig{figure=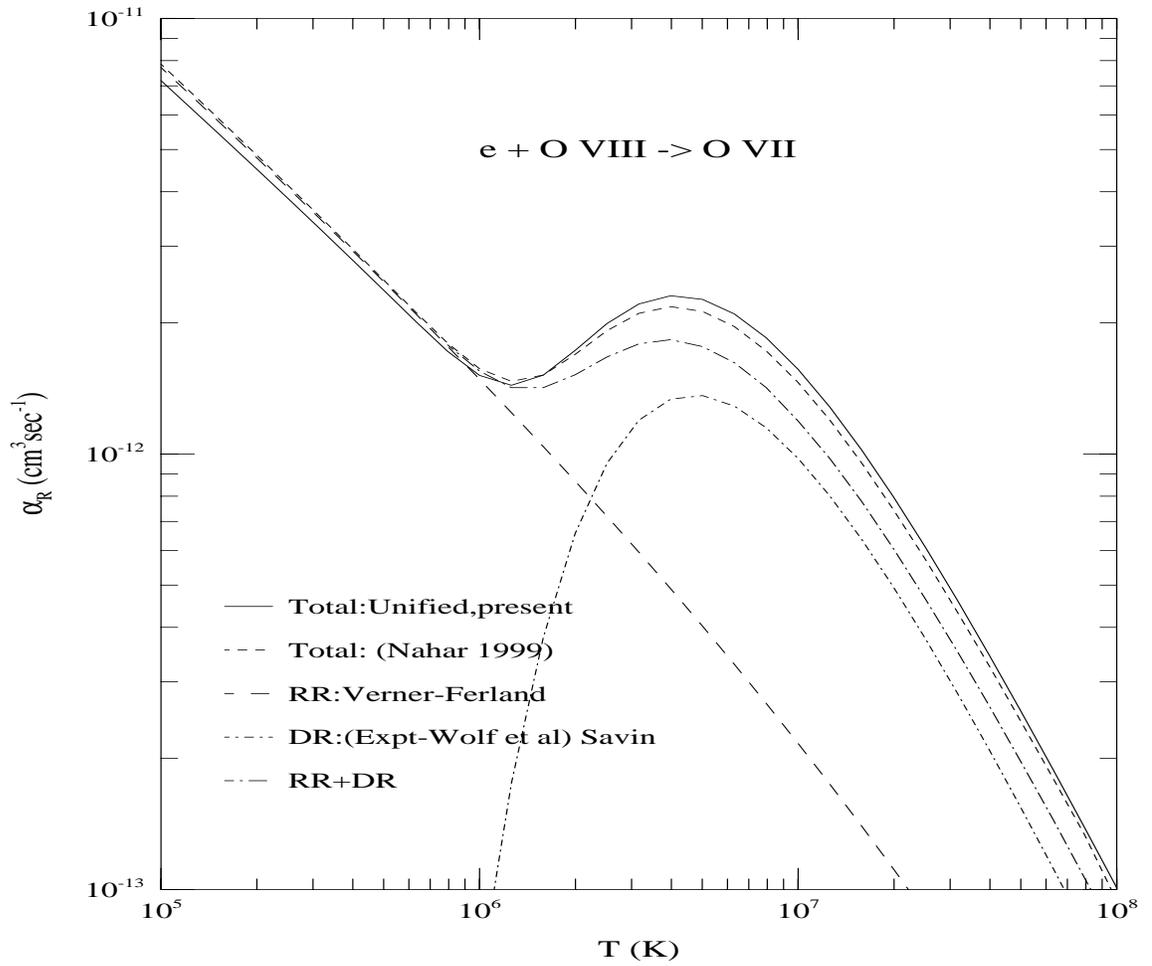,height=15.0cm,width=18.0cm}
\caption{Total unified recombination rate coefficients for O~VII: present
BPRM (solid); total (dotted, Nahar 1999), RR rates by Verner and Ferland
(1996) (dahsed), DR rates by Savin (1999) from measured cross sections
by Wolf et al. (1991) (dot-dashed).}
\end{figure}

\begin{figure}
\centering
\psfig{figure=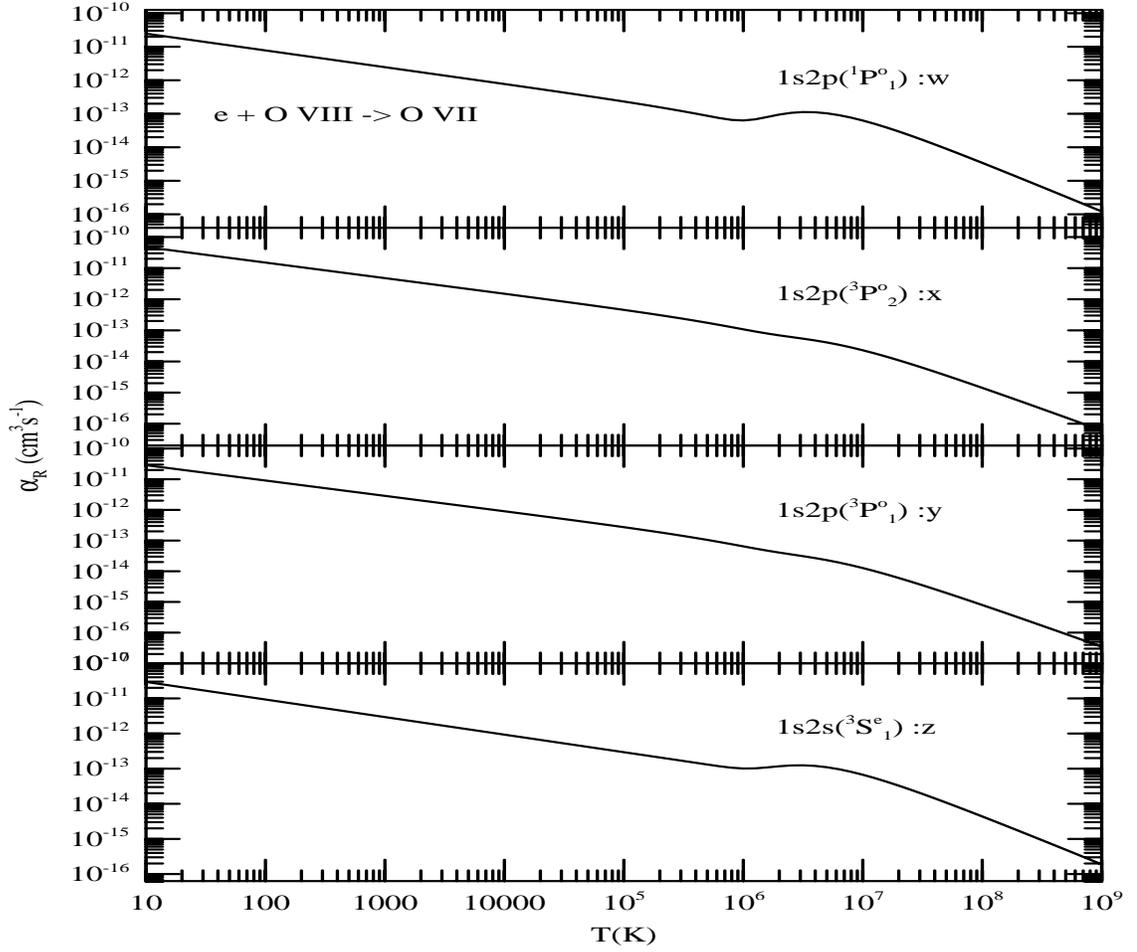,height=15.0cm,width=18.0cm}
\caption{Level-specific recombination rate coefficients for O~VII into
the excited  n = 2 levels responsible for the prominent X-ray lines w,
x, y, and z.}
\end{figure}

\end{document}